# Trust architectures in payment systems: the great bifurcation[i]

D . Boullier*, N. Sivakumar**, M. Crepel**, S. Juguet***

*EPFL, Lausanne (Digital Humanities Institute), **Sciences Po Paris (Médialab),

***Agence What Time Is It, Paris

January 2017

Abstract:

Payments architectures are on the verge of a great bifurcation that must be documented in order to be debated. Google is moving towards a quasi bank while Apple and Google disseminate payment systems over smartphones. At the same time, block chain might become a distributed ledger introducing a radical new model of trusted third-party. The detailed history of credit card systems helps understand why the game of security has always been trigged by a delegation process of the risk to third parties and by the cat-and-mouse game of security and fraud. Technologies were designed to solve these issues but have always been closely related to innovations in institutional assemblages. These payments systems shape our social life and the stakes of trust that we put in these architectures require a truly political examination.

**Contents**

Introduction

Conclusion

**Introduction**

When, in 2014, Google sought an agreement from the Federal Deposit Insurance Corporation, one should have known that the online payment system was doomed to change radically in the next few years. However, history has taught us that payment systems have never actually been as stable as they might have seemed to the laypeople using the systems. All changes have been focused on building trust through two major features of these systems : the trusted third party and the security of exchanges. But this move from Google takes place in a very specific moment, when trust is no longer guaranteed in digital networks, due to the predatory uses of personal data traces from internet platforms and to extensive wiring of the network by intelligence agencies, as evidenced by the Snowden leaks. These weaknesses of some major supposedly trusted third parties became more and more visible through the tremendous extension of data breaches of various kinds, from Sony to IRS, Yahoo and dating sites, in the recent years.[ii]

The post-Snowden mood has put pressure on digital architectures that has not yet been materially affected by the lofty promises of privacy-by-design. Payment systems cannot take for granted the safety of any feature of their underlying architecture. Getting back to the history of credit card systems provides an opportunity to observe similar situations where security flaws were addressed by moves for more secure transmission techniques and more trustful third parties. The move made by Google mixes two types of evolution: becoming a quasibank by relying on state insurance for its payment activity and using extensively the tokenization system which is becoming the standard for online payments, while empowering the fluidity of Android Pay (launched in September 2015) from the end-user point of view. However these improvements are made with a cost: the full and extended centralization of personal and financial data into the hands of Google servers. As it is often presented, this is a very user friendly move and an almost inescapable one, since code has its laws (to the point that « code is law » according to Lessig, 1999) that no one could underestimate. However, there is precisely at the same time, another architecture that appears to fulfill the conditions for becoming a challenger of this centralized payment system. The blockchain system as implemented for validating Bitcoin currency operations and transactions, may well be adapted for generating a distributed trust system. Specific variations of the blockchain-based processing systems, like one proposed at MIT (Enigma by Zyskin, Nathan and Pentland, 2015) extend the basic properties of a decentralized database to protect the privacy of transactions through a novel technique known as "secure multi-party computation". The changes that can occur in payment systems such as the adoption of the Google system are of huge consequence for issues of sovereignty. They should be handled very carefully by all authorities of control, and the availability (to be validated) of other distributed frameworks should be considered as a good opportunity to politically discuss the choices of architecture that are often processed as « subpolitics » (Beck).

This paper will follow the main milestones in the history of credit card systems. Each type of architecture (closed-loop, open-loop systems) will be explained as the design of a technical and institutional distribution of risk. The important changes in the level of risk with the online payments will be investigated through a number of new intermediaries in charge of alleviating the burden of the risk. However, the introduction of Google and Apple within this game is described as a significant move towards centralization. At the same time, however, the blockchain distributed framework may become an alternative resource that has to be carefully designed in order to play a real challenging role.

**Section 1 History : a cat-and-mouse game reaching a tipping point**

The technical foundations of the credit card actually predate the invention of the cards themselves and instead lie with the development of a widespread cheque clearing in the United States in the early twentieth century. A networked system of clearinghouses needed to be developed in order to foster the use of checks as a method of payment that could be used at different merchants who banked with different institutions. Otherwise, cheques were instruments that could only effectively be used between payers and payees that shared a banking institution. While the idea of a bankers' clearinghouse was not novel, with earlier examples dating back to the late eighteenth century, what was unique in the United States at the time was the co-development of both the clearinghouses and a the US Federal Reserve Bank as as a central institution that could regulate the clearinghouses. The clearinghouses themselves did not entirely solve the problem of efficient cheque usage. In fact, the clearinghouses introduced an increased set of transaction costs and inefficiencies for the processing of cheques as several competing clearinghouses were established. The Federal Reserve ultimately intervened, setting up a national check clearing system that was funded by taxes and free to use for members. While this infrastructure itself would evolve and not directly form a base for a universal credit card system, the effect of the development of a national clearinghouse in 1915 was one of the first developments that created an expectation for universally accepted payment technologies that would operate across different banks and merchants but also be trustworthy and resistant to fraud. During this historical period, the challenge always consisted of inventing new trusted third parties that could handle the technical complexities of a new era. Sometimes, the technology served as a trigger to form this third party, sometimes, the fraud requested a legal intervention beyond the technological realm. This is why, until recently, states were at some point always called on for assistance. And still are when Google comes to ask the insurance of the Federal Deposit Insurance Corporation. This tells a lot about how the most innovative and profitable technical developments do not manage to be their own legal and trustful reference, although the giant platforms are coming close to this sovereignty. However, a precise tracking of the history of Credit Card and payment systems institution will show how delicate is the design of such a status.

The figure below illustrates the main landmarks in this history even though we cannot get in details in this paper. It must be reminded that technical moves on the consumer side (e.g. from imprinter to magnetic strip) were designed for fraud reduction purposes and in the end lasted only a few years until weaknesses were exploited to a large scale. Some recent breaches managed to get into the very production chain of cards such as the SIM (subscriber identity module) card authentication keys stolen in 2010 by GCHQ and NSA intelligence services from Gemalto, the Dutch corporation that is the largest producer of SIM cards in the world. The same fraud-driven innovation scheme is true for institutions and infrastructures. In 1973, the creation of SWIFT (Society For Worldwide Interbank Financial Telecommunications) was a critical step in standardizing exchanges between the 3000 financial institutions that participated. It should have been the safest place in the digital financial system. However, in April 2016, Swift warned of fraud cases in its system. Such a trusted third party itself might well be under attack and this could trigger, in a next hacking scenario, a radical loss of trust that would affect the whole economy. It seems quite clear for security experts, that the design of the network (from internet distributed principles to the increasing centralization around some nodes and to the most critical specialized networks such as Swift) did not pay sufficient attention to security for the benefit of speed, which is a key value in financial and opinion economies (Orlean, 2011) and for ease of use. The data breaches rhythm seems to increase and their consequences proportionally aggravate. In this context, a deeper understanding of what is framing our everyday life, i.e. the worldwide payment infrastructure, is crucial, especially when new entrants such as Google may pretend to technically relieve our stress, in a now famous solutionist stance (Morozov, 2013).

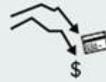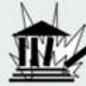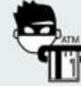
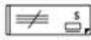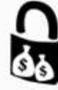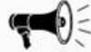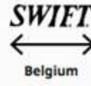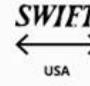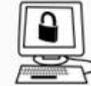
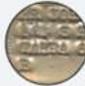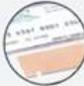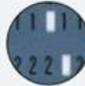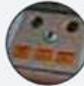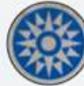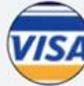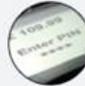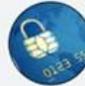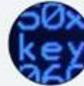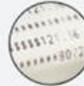

The early charge cards were issued by department stores, the oil industry and the air travel industry. Credit cards allowed the immediate extension of short term credit to buyers fostering loyalty as well as larger impulse purchases. These cards often took the physical form of a metal plate or fob that could be attached to a keyring and would contain the identity of the user, an address, the name of the issuer and an identifying number. Card purchases were first processed by hand, with merchants manually copying the information from cards during each transaction. This process proved to be highly error-prone and unreliable, causing losses to merchants. This limitation will be one of the constant incentives for technical innovation: reduce the role of humans who are least reliable link in the chain. Technical systems are always made of these chains of translations (Callon, 1986) from one state of the information to the others and the number and the quality of each of these links must be checked constantly, generating this maintenance and control work at each articulation (Strauss, 1978). But the technical solution is often favored even though it creates some new opportunities for failure

The first major technological change to credit card processing came in the form of embossed cards that could be physically copied using carbon paper. This system greatly reduced copy errors from manual processing and materially reduced related losses. However, credit cards were used quite infrequently at the outset. The oil industry decided in the 1930's that one technique for fostering widespread adoption of credit cards was to simply distribute cards en masse to the public for free. As Americans began to purchase more and more automobiles, the oil industry's technique worked as credit cards provided an easy way for drivers to refuel their vehicles. However, the oil industry quickly came to realize that as adoption of credit cards began to increase rapidly, so did opportunities for fraud, setting in motion a technological cat-and-mouse game that continues to the present to develop new mechanisms for minimizing risk and fraud of credit card usage.

We cannot get into more details about each step in the design of credit card systems**.** The following thick description of the various payments architectures in the credit card environment will document this race for more reliable third parties and technical systems at the same time. The rule seems to be always that of adding new layers of security until encryption and more and more delegating the risk to trusted third parties. Delegation is a key process in technical design and can be a relevant concept for both institutional and technical systems. However it comes to a cost, the articulation costs, well described by Anselm Strauss.

**Section 2 Payments architectures in Credit Card systems: a story of delegations**

**21/ Closed-Loop Payment Systems**

According to Benson and Loftesness (2010), payment systems can be primarily structured in two ways: either "closed-loop" or "open-loop." Different credit card systems have adopted both models for various historical reasons. These two early models of closed-loop systems of payment also come from the world of credit cards, as both American Express and Discover established closed-loop payment systems (in 1958 and 1949, respectively) (Steam, 2011). Closed-loop systems require direct contact between all entities and the payment system: both the merchant and the client must be registered within the payment system of Amex for instance. These systems are much easier to regulate than open-loop systems as rules are set only by the payment system and directly promulgated to the their subscribers, unlike in the case of an open-loop system where rules are layered by each intermediary institution and slow to propagate through a given network. The delegation is minimal in this model and the costs of articulations are reduced but the responsibility is complete. Closed-loop systems have historically been very slow to grow due to the high initial cost of establishing relationships with end users. As a result, many credit card companies are moving to an open-loop model even if they were historically established as closed-loop systems.

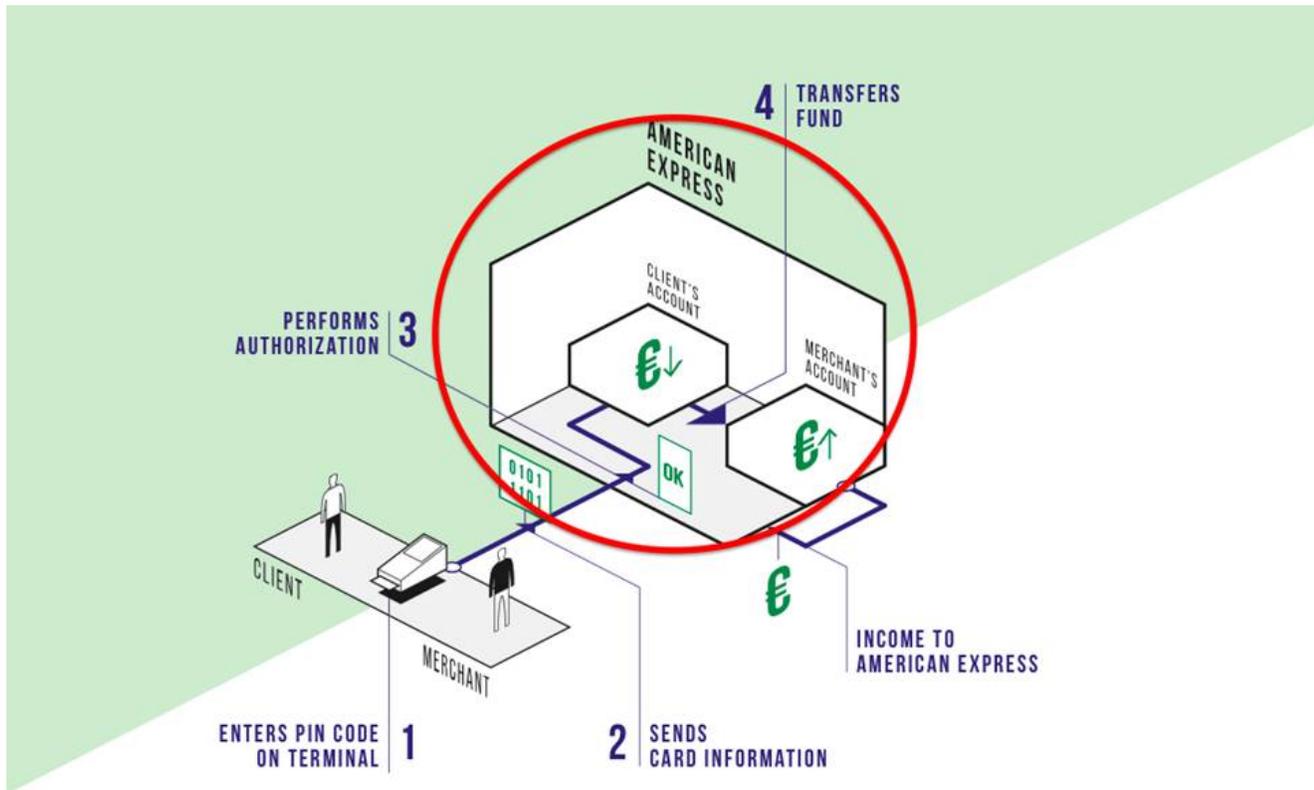

**CLOSED-LOOP PAYMENT SYSTEM
CARD PRESENT PURCHASE** | **SCENARIO 1**
PAYMENT SYSTEM

## 22/ Open-Loop Payment Systems

An open-loop system is a payment system that operates on a variety of intermediaries to subscribe to a payment system. These intermediaries in turn form relationships with other parties in order to form a distributed network that is coordinated indirectly through the payment system which functions as a hub. There are several trade-offs to this approach to structuring a payment system. Transaction costs are increased greatly and the efficiency of the system is hampered by an open-loop structure. However, the scalability of the system is much greater in this approach versus one that requires direct coordination between parties. The trade-off between security and fluidity (through a standardized and extended payment system) is still alive and well, and cannot be solved in an ideal solution, although innovators pretend to reach this stage for every new assemblage (Latour, 2005). Furthermore, over time, open-loop networks have become supported by a relatively predictable system of liability and risk assignment which has become a primary motivating factor for continuing to foster this form of organization even when technological developments have now made it possible to perform trivial bilateral transactions. This is important from the perspective of insuring against the risk of damage, as the open-loop model of payment systems helps to modularize and distribute the cost of insurance across a variety of different institutions with varying capabilities of accommodating risk.

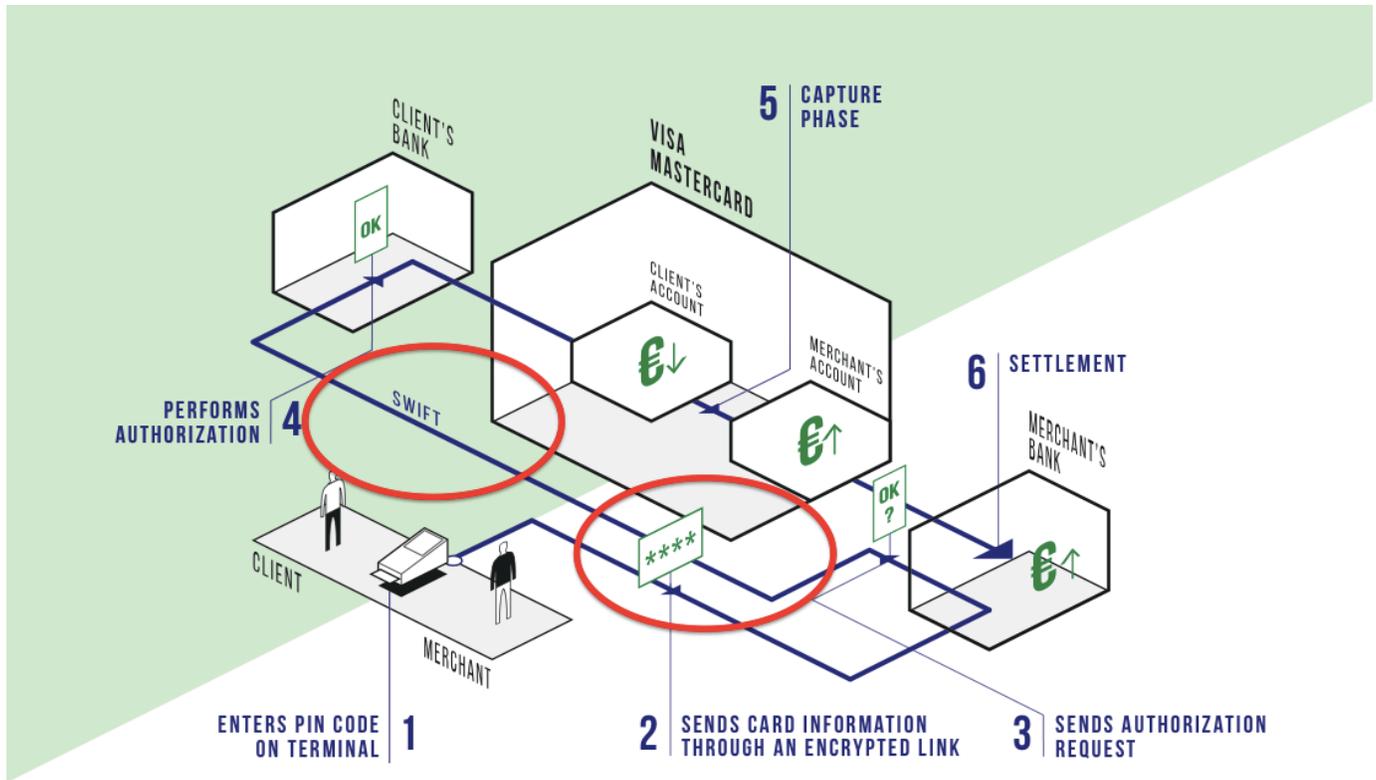

Card networks, at least in the context of physical "card-present" purchases, are designed to mitigate some of the risks of fraud associated with other payment systems. This is done by the implementation of a card network verification system that operates in parallel with the fund transfer clearance and settlement mechanism. Credit card transactions are subject to a pre-authorization transmission where the actual transfer of funds is preceded by a request to the card network that verifies whether actual funds or credit balance is available to cover a given transaction. The transaction only proceeds to the settlement phase if an authorization is received. This is the next step into a design of redundancy for control purposes that create friction in the transactions at every step of articulation but at the same time build trust and delegate or distribute the risk among the stakeholders. Importantly, from a risk management and insurance perspective, card networks explicitly indicate that the reception of an authorization message transfers the liability for fraud risk away from intermediaries to the card network. This means that intermediaries like merchants need not insure themselves against this kind of credit card fraud and that the risk becomes aggregated in the central card network, even in the case of an open-loop structure. The importance of aggregating risk is that what is known as "brand" in the technical parlance of payment systems and is considered to be an important driver in supporting the growth of credit cards by assuring users they will not be held responsible for the risk of fraud when using credit cards. This specific notion of "brand" differs from both the common usage of the term as well as the legal usage of the term that is rooted in trademark law. Here "brand" refers not only to a specific brand in the sense of a given card network like Visa or MasterCard, but also to the payment mechanism itself where "credit card" is seen as a different brand than "cheques" for example.

### 23/ Regulation of Credit Card Systems

The regulation of credit card payment systems is complex and has evolved overtime. It must be told more precisely since the institutional design is as much critical as the technical one in security matters. Visa and Mastercard both were started as systems owned by banks while American Express and Discover were started as non-bank corporations. Threeof the major credit card networks now operate as non-bank entities, while American Express received approval in 2008 to be considered as a "bank holding company". This distinction is important because the laws that regulate banks and non-bank institutions are distinct, both in the US and in Europe and it helps appreciate the Google move towards becoming registered as a bank as we mentioned in our introduction. This is particularly important with respect to insurance, as entities defined as banks, either explicitly or de facto, often are under legal mandates that require that they maintain insurance. Examples of these regulations can be seen in the US Federal Deposit Insurance Corporation (FDIC) which is mandated by law to insure bank deposits and variety of individual member state laws that have been implemented in Europe pursuant to the passage of Directive 94/19/EC such as the French Fonds de Garantie des Dépôts (FDG). Instead, as credit card networks now operate as non-bank institutions, they are governed by a combination of the networks' own rules and mixture of legal regulation that is specifically targeted at credit card systems. However, it is important to note that rules targeted specifically at credit cards are more recent than banking regulations, which tend to date to the Great Depression, and thus are less tested than banking regulations. Legal regulation of credit cards also tends to be less explicit and forceful than banking regulations, with marked examples of this being seen both in the 2009 Credit CARD Act in the United States and the 97/489/EC: Commission Recommendation of 30 July 1997 concerning transactions by electronic payment instruments and in particular the relationship between issuer and holder in the European Union.

While the comparatively lax regulation of credit cards may appear to position the system as one that is inherently more risky and subject to fraud than traditional bank-based payment systems, credit card networks have developed in unique ways in comparison to other payment systems that allows them to internally mitigate and manage risk. Most current bank-based payment systems are described as "thin systems" that are focused on maximum efficiency and the reduction of transaction costs associated with payments. Credit card networks, on the other hand, have developed to be "thick systems" that are heavily resourced in such a way that they mobilize their own resources to manage risk, often acting themselves as insurers against fraud on behalf of their intermediaries and end users. Benson and Loftesness hypothesize that this is because credit card networks are motivated by the generation of profit rather than the reduction of costs, thus having more access to capital that they employ on their own volition without formal mandates in order to further secure future profit by effectively managing risk. Furthermore, card networks compete among each other for exclusive intermediary relationships and for end users, resulting in relatively high competitive pressure to produce rules that encourage intermediaries and end users to mitigate risk but are not overly burdensome and thus ultimately positioning card networks to structure their operations in such a way that prioritizes risk mitigation.

### Section 3 Intermediary Risk and online payments

In the case of open-loop credit card networks like Visa and Mastercard, intermediaries still continue to bear some significant risk. Intermediary card issuers, for example, bear the risk of credit exposure and related fraud. Issuers, who are almost always banks in the US and Europe, manage this risk through their recourse to public and private tools that model and evaluate credit risk exposure, not only for credit cards, but for all the different credit-based activities of the given bank. Issuers are also in charge of managing chargebacks in transactions where an end user disputes a given credit card transaction. In some cases, pursuant to rules set by the card network, an issuer can reverse a credit to a merchant's account in eligible disputed transactions shifting the risk of purchasing fraud away from the issuer and card network to the merchant. Card issuers can also shift risk to end users pursuant to card network rules, but often position themselves to accept this risk again through the issuance of insurance products directly to end users. In a recursive fashion, card networks have now also established rules and policies for sharing insurance risk with issuers, creating a situation where insurance forms a unique tool that is used to manage a kind of "last mile" risk that directly affects end users. Even though end users are responsible for a portion of the risk through the subscription fees they pay annually, this seems to be often forgotten, especially when they experience successful dispute resolution with merchants thanks to friction-less intervention of credit card networks. In this way the complex design of a distributed risk architecture becomes a black box, but one that is adorned in Christmas wrapping paper! This is how credit card networks as open-loop systems gain reputation and extend the realm of their activity. It would have been expected to find the same simplicity and transparency in an online payment system where the burden of physical cards and terminals could have disappeared in an intangible exchange system. However, this is not the case as every end user has likely experienced.

Important changes in the level of risk of payment appeared with the extension of online payments, a context in which a physical card is no longer present in the terminal of a merchant. The chain is extended for the benefit of merchants and clients who came to adopt online commerce as a part of daily life, even though the process took over two decades to become familiar to a large number of clients. But this extension and the fluidity that it brought for immediate transactions generated a new generation of fraud. Outside of credit risk, the major categories of risk for credit cards are associated with fraud such as the theft of cards, counterfeiting, intercepted cards, identity theft and creation, unauthorized use and PIN fraud. The risk of non-credit associated fraud has been historically low in the context of card-present transactions, but has become much more pronounced since the popularization of e-commerce which has created a norm of card-not-present transactions. Mitigating fraud in the online context has required card networks to move beyond only their own private regulations and supporting legal regulations and to turn to a variety of technological measures to mitigate risk. These include the promulgation of best practices that relate to data protection practices, hardware implementation at point of sale, requiring cryptography for card-not-present transactions, developing new physical technologies that augment magnetic stripes with "smart chips", and also the development of new protocols, like tokenization, that limit the exposure of credit card data in online transactions.

OPEN-LOOP PAYMENT SYSTEM CARD NOT PRESENT PURCHASE | SCENARIO 3 PAYMENT SYSTEM

ÉTUDE : SCIENCES PO. - MÉDIALAB | CONCEPTION GRAPHIQUE : WHAT TIME IS I.T.

From the client experience point of view, e-commerce seems to require a constant extension of barriers, that have been added along the years, in order to associate with a sufficient level of trust the body of the client and the flow of information within the system. The CVV (Card Verification Value or Card Security Code) must be read optically on the other side of the card and can prevent the capture of numbers from non material opportunities (credit card number in a mail as in phishing operations) or from basic card image capture in ATM for instance. Anybody, somebody must be present, to operate it at the right moment, despite the intangibility of this world. This simple printed number (CVV) seems weak compared to digital methods but it reintroduces the materiality of the card as a reliable and more controllable feature of the system. The same is true for the body that is one of the inescapable ends in the chain of trust (the other one being the supreme guarantor: the state). However, this simple artifact, the CVV, was insufficiently safe, and banks (and not the credit card issuers) now deliver a CAP (Chip Authentication Program) device and oblige their client to dial their Personal Identification Number on small devices equipped with a complex calculation routine that generates tokens with a unique validity. The card must be present but the materiality of another terminal is introduced and associated with a cryptographic technique.

In addition to these technological measures, Visa and Mastercard, who are open-loop networks, have developed a novel system of "master merchants" where they rely on aggregate intermediaries like PayPal and Square to implement a hybrid open/closed model that allows them to continue to benefit from the strengths of an open-loop structure while leveraging the ability to more efficiently regulate closed-loop systems in the online context where risk exposure is greater.

The highly vulnerable online context leads to a new layer of complexity for not only the user but for the whole chain as well. Specialized online payment service providers, like PayPal, have structured themselves as closed-loop systems that operate on top of open-loop systems, such as by piggybacking on credit card systems. Closed-loop systems treat transactions on an individual basis and thus treat risk as a discrete component of each given transaction unlike in an open-loop system where risk is often aggregated in intermediaries through batch-processing and settling of transactions.

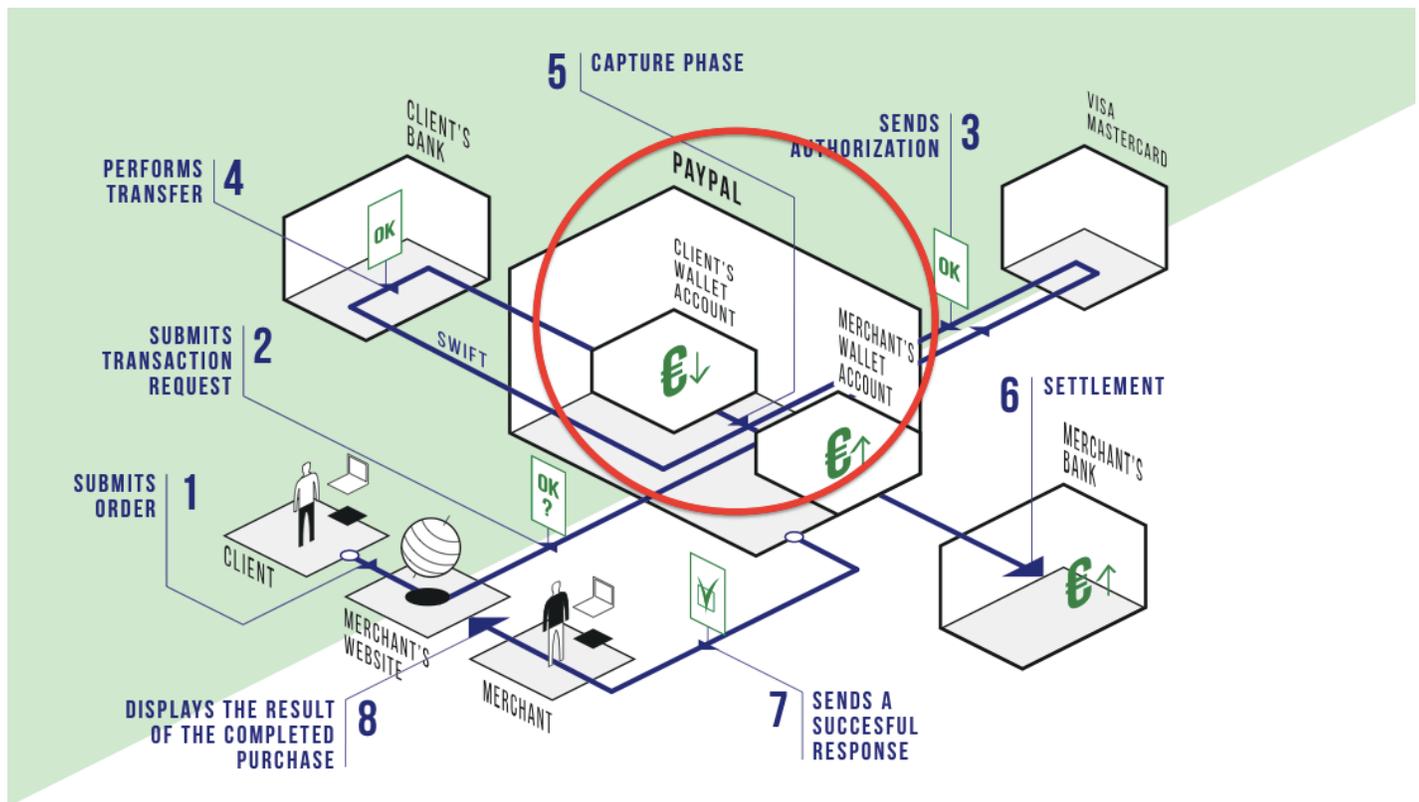

OPEN-LOOP PAYMENT SYSTEM CARD NOT PRESENT MASTER MERCHANT PURCHASE | SCENARIO 4 / PHASE 2 PAYMENT SYSTEM

**Section 4 Special focus on token vaults and Google and Apple pay systems. Why terminals design is not the main issue**

The next step in improving the control of transactions relies on tokens which validity is limited to one specific transaction. ANSI's[iii] X9 group defines it as « a surrogate value used in place of an underlying sensitive value (USV) in certain, well defined situations but not in every way that the USV is used ». The « real-time » exchange builds a relationship that is supposed to be more akin to the face to face transaction, since nobody can store and reuse the tokens generated by a secure system. These disposable tokens, however, must be connected to a valid account through some mechanism and this is accomplished through the establishment of a "token vault."  Our emphasis on this « token vault » stresses the need for a shift from « user-friendly » terminals storytelling that purveyors of new technologies are trying to spread to the changes in infrastructure that are seldom decoded for the layman.

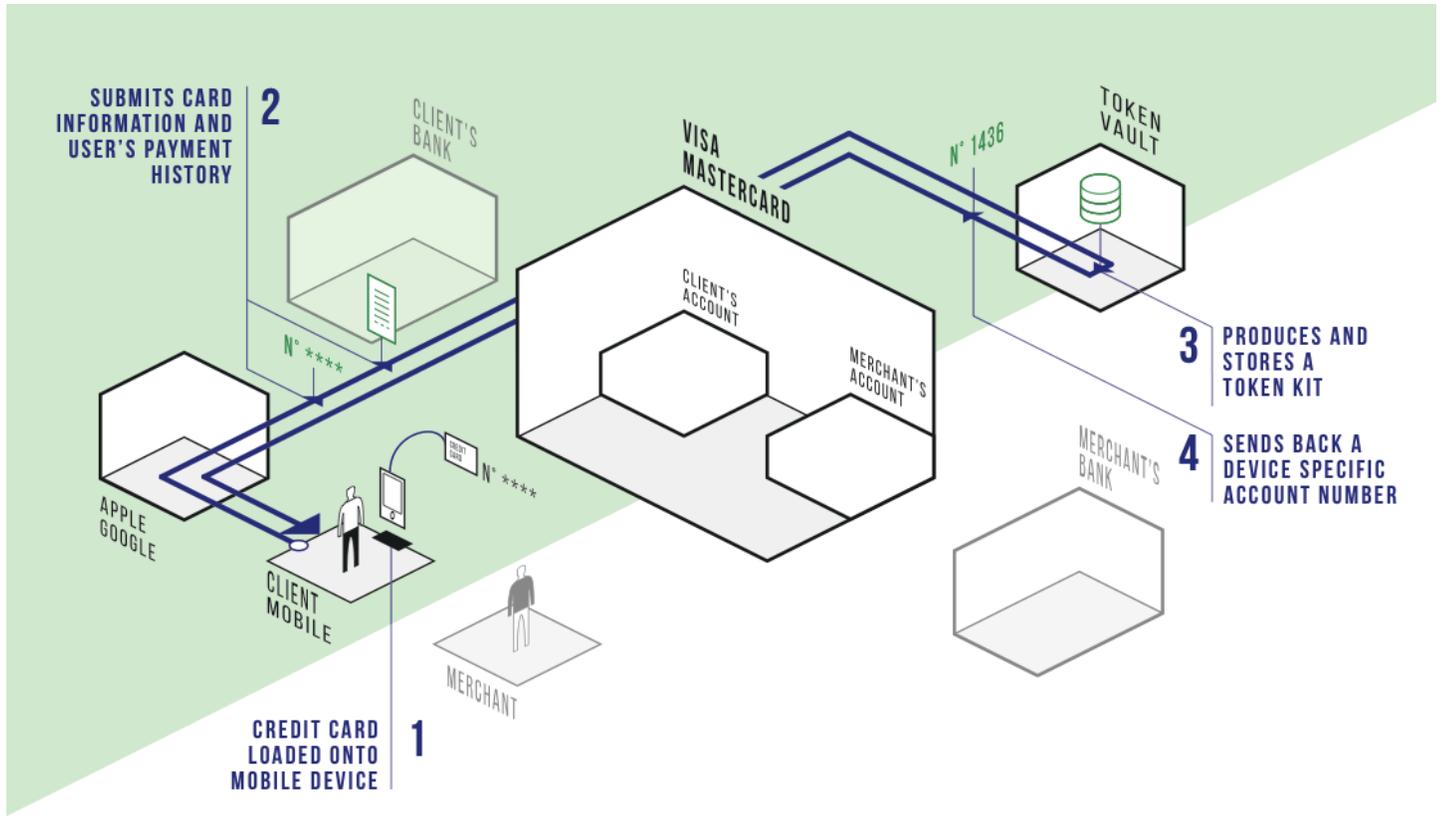

**OPEN-LOOP PAYMENT SYSTEM DEVICE PRESENT PURCHASE WITH RFID AND TOKENISATION** | **SCENARIO 5 / PHASE 1** PAYMENT SYSTEM

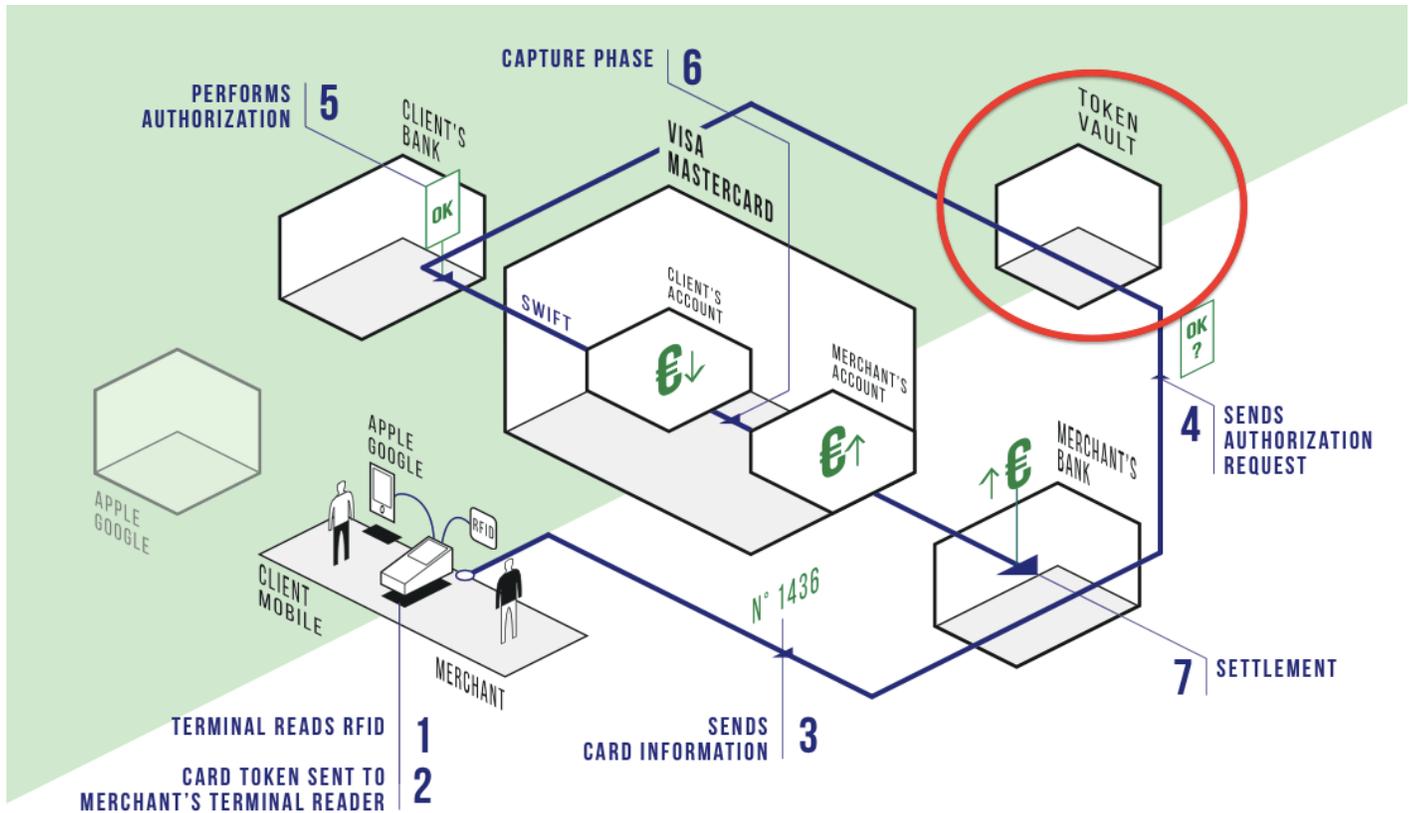

**OPEN-LOOP PAYMENT SYSTEM DEVICE PRESENT PURCHASE WITH RFID AND TOKENISATION** | **SCENARIO 5 / PHASE 2** PAYMENT SYSTEM

Thus, another technical barrier is settled and along with it, a new organizational layer is implemented, in the form of the « vault ». This is how the general architecture becomes increasingly complex, as any new technical layer leads to the introduction of a new stakeholder. Risk allocation among links in the chain must then be revised and clearly re-defined. The token vault architecture is one on which Google and Apple take shelter when introducing their new payment systems. The choice of tokenization vs. encryption is justified by the need to reduce auditing costs and the impact of the workflow. However, the debate about encryption will still be alive for long.

The focus often put on the simplified interaction between the user and the terminal can be reframed in this context as a tighter "assemblage" between the body and the machine. As usual in this search for guarantees, the role played by the body's recognition, either the fingerprint or the face, is emphasized, because it adds a new level in the reliability of the link « body » in the transaction chain. This is not only spectacular and supposedly simple (not so much as to handle a banknote, however), but it builds trust upon the traditional legal modern reference. Identity is supposed to be unique from the individual point of view and the body encapsulates this proof. All moves in authentication procedures were driven by the search of the final unique proof of identity: the description made by civil servants in the first age of ID in order to describe the physical properties of the registered citizen, the invention of the fingerprints and their database as a reliable source for criminal records (Bertillon), and the now routinized DNA authentication in scientific police activity. The legal artefact of identity is reduced to the biological proof of unicity, even though science fiction movies such as *Gattaca* anticipated the possible frauds of these sophisticated tools. The dematerialized world and the constant reduction of everything to digital traces reach a limit when it comes to certify IDs and to build trust. It requires a move towards the analogical body, even though algorithms can transform its components (fingerprints, gaze, face, DNA) into computable material.

But at the other end of the chain, the open culture of connectivity, the disruption brought by new entrants in all markets that can act as platform and third parties, manipulating terabytes of data and personal information, reach another limit: they need to rely on a indisputable third party. The token vault is just one more move to design a third party with the minimal engagement, the minimal responsibility but still able to act as a true intermediary.

The trend is clear: always try to delegate these security processes to someone else i.e. do your best to pass the buck. This is not a mere technical complication but a delegation encompassing all the stakes of transactional validation. But as Google's move for the State guarantee demonstrated, there is no limitation for such a delegation: these new stakeholders do not shelter enough in the end and this is why the state is still the last guarantor. There is some kind of irony in the fact that the authority which is looked for comes to be always this outdated « State », even for those powerful platforms which enter the market of online payments. The body and the State remain the two sources of trust in these operations as in any other security process (Boullier and al., 2007): this does not look so innovative, does it? However, new insights provided by the technology of the block chain may be able to shuffle all these traditional "assemblages" through its distributed architecture.

**Section 5 The great bifurcation: centralized or distributed systems of payment**

Block chain-based payment systems may become an alternative mechanism that may mitigate some risks of credit card systems, be they open or closed-loop, while presenting their own set of risks that must be contended with. The block chain is closely associated with the crypto currency known as Bitcoin as the technologies were proposed together in a paper written by (Satoshi Nakamoto, 2008). However, although the two technologies are often considered to be parts of a single system, the block chain is in fact an independent cryptographic public ledger that can be used in a variety of different computing systems including other virtual currencies known as "altcoins" or even for other general purpose computing systems.

The block chain is a distributed database designed to keep track of the spending of Bitcoins. In particular, the system is intended to prevent double-spending of currency by implementing a securely time-stamped chain of transactions in an expanding data structure that allows the verification of transactions through time. The code for the block chain is distributed under a permissive free software license. The verification of transactions is performed through hard computational challenges that are solved by the computing nodes that are constituents of the block chain network that in turn also maintains its ledger spread across the network itself. In this way the transaction and clearing mechanism of the network does not rely on a central entity like a bank to function instead relying recursively on its own participants. In the case of a virtual currency like Bitcoin, participation in the distributed network is incentivized by providing a portion of each transaction's value to the participants that contribute resources to solve the challenges that verify the given transaction.

As seen in block chain-based technologies such as Ethereum's Etherchain, IBM's Open Blockchain, MIT's Enigma and the Hyperledger Project which is backed by several major financial institutions, the fundamentally distributed ledger technology of the block chain is being leveraged for a variety of financial technologies that extend beyond that of virtual currencies including credit card-like payment systems that can be used by consumers. This decoupling of virtual currencies from the underlying technology means that a payment system can bypass the social encumbrances of new currency, such as those noted by financial sociologist Nigel Dodd, and bootstrap technological uptake on the familiarity of card payment systems (Dodd, 2014).

Relying on block chain technology for card transaction processing allows for the mitigation of several of the risks that we have noted with credit card systems. The decentralized nature of the block chain means here that the responsibility for storing and processing payment data is removed from the hands of the stakeholders and placed instead on distributed computer code in a Lessigean re-allocation of risk. The architecture of the block chain and in turn of the payment system that is built on the block chain sets the security and privacy requirements by which participants in the payment system must abide if they are to participate in the system at all. This in turn reduces the reliance on an undisputable third-party for maintaining a secure vault or even from needing to audit participants in the system as they must in a traditional credit card payment system.

The reliance on a block chain, however, is not a panacea and presents its own potential points of failure and risk. The first is that the code for the block chain itself must be sufficiently robust to actually provide the guarantees that it claims. As computer scientist Steve Bellovin (2015) has noted, a single programming bug or error can theoretically be a major security flaw in software and this applies no less to a technology like block chain. While block chain is licensed as free software and subject to the practices of public open source code review methods, other critical free software has succumbed to major security flaws in the recent past. Next, the payment system that is built on the block chain must again function as intended and must be designed with the correct parameters to ensure mitigation of risk. The block chain itself is merely a ledger and does not itself provide any guarantees beyond the correctness of the transaction that it records. New systems like MIT's Enigma propose a privacy-sensitive network design that leverages the block chain to support a system that provides verifiable but secret transactions[iv]. However, these technologies are nascent and have yet to be subjected to rigorous testing to ensure their security and privacy preserving features.

Finally, the materiality and functionality of the block chain itself can subject these systems again to the same political and social forces that affect more traditional payment systems. While the block chain is often discussed in a conceptual abstract, the software must run on physical computers that are administered and owned by human agents. This means that the incentives for joining the system must be carefully engineered to induce participation. The efficiency and functionality of the system is also reliant on the quality of the participants in the network. As recently illustrated by the forking of Bitcoin into Bitcoin Core and Bitcoin Classic, these practical issues can have a serious effect on the technical architecture and functioning of a block chain system. Briefly, the forking of Bitcoin was the result of a controversy over whether to increase the block size of each block in the block chain. While seemingly an innocuous change to the block chain, the effects of this tweak could be wide ranging.

Maintaining a smaller block size means that the computing resources required to participate in the "mining" process of transactions are reduced but at the cost of network performance. This means that a more heterogeneous network would result in slower processing time for each transaction on the network. Proponents of a larger block size argue that this change would facilitate faster transaction times and more efficient use of the block chain for a variety of computing tasks beyond a simple currency that could require more information to be maintained for each transaction. However, the increased resource requirement to contribute to such an expanded block chain could alter the barrier of entry to the network such that only those with great resources, like states, large corporations, or banks, could participate in the network. This would in turn undo the benefits of relying on a decentralized network for payment processing by effectively re-centralizing the block chain the hands of a small set of actors.

Furthermore, effective participation in a block chain becomes limited to specialized hardware as the resources required for "mining" increase. Processing of block chain calculations is done much more efficiently on application-specific integrated circuits (ASICS) than on consumer-grade hardware such as personal computers, graphical processing units or field-programmable gate arrays (FPGAs). Since the production of ASICs are limited to specialized fabrication centers in very few places in the world, practical access to participation in a resource-intensive block chain could be further limited by whether access is granted to the necessary hardware by the producers or states that house the producers. In this way, the potential of the block chain for a secure, decentralized payment system itself produces a paradox insofar as maintaining the speed and efficiency of current payment architectures could require a block chain to adopt the same limitations that it is supposed to alleviate.

**Conclusion**

As the story of the block chain tells us, there is no such a thing as a frictionless transaction. In payment systems, the question remains of the best allocation of resources to secure data and the best distribution of risk among stakeholders. Even the most distributed model faces the same challenges of optimization of speed *vs* security. However, the technical opportunity offered by the block chain is crucial to avoid being trapped in the fatal trend towards more centralization in the hands of the oligopolistic platforms that dominate internet. There is some room for competing architectures provided that traditional stakeholders like banks or credit card networks understand the need for changing their principles when faced to the powerful move of Google (or Apple). Payments systems have a long and sensitive history that had to be told in order to remember the materiality of their architecture and its tight coupling with legal procedures to mitigate the risk. The trust that is built in the process gets down to every individual consumer and the loss of confidence that could propagated from mind to mind at high speed (Tarde, 1969) if some breach could let believe that no one is in charge to ultimately cover the risk. Our information systems were designed in the last 40 years with the constant privilege for speed over security. This was not the case for the credit card system that was used to the cat-and-mouse game with fraud. However, innovation speed and computing power increases put these architectures in trouble and not able to keep the pace. A paradigm shift is ready to take place: will platforms become the last and final guarantor in financial transactions as Facebook does for ID management in many third party applications and services, thus adopting a very centralized approach and disrupting all financial institutions and services? Or will these trusted institutions be able to invent the new distributed payment architecture that will preserve their role by completely changing the allocation of trust in favor of distributed systems such as one afforded by a block chain? The careful design of each ring in the chain of security still remains a key condition of felicity beyond the mere adoption of principles. "Conventions" require a long-term and open effort to be built. However, the clock is ticking, since the constant expansion of data breaches might well oblige everybody to react under the pressure of a « Fukushima of Internet » that would mean a general collapse of trust.

**About the authors**

D . Boullier, Professor of Sociology, Digital Humanities Institute, EPFL, Lausanne, Switzerland,

N. Sivakumar, Computer scientist and lawyer, médialab Sciences Po Paris,

M. Crepel, Sociologist, Medialab Sciences Po Paris,

S. Juguet, Designer, What Time Is It? Paris

**Notes**

[i] This paper is based on an historical investigation of credit card systems, mostly done by N. Sivakumar, as a part of a project at Sciences Po, Paris, medialab, funded by Axa Research Fund « Insurance for Building Trust and Enabling Big Data » (2014-2016). The definition of the framework and the direction of the project were in charge of D. Boullier. Its management was done by M. Crepel. The visual displays of payment systems were designed by Stéphane Juguet and his team (What time is it?).

[ii] For a visual watch of data breaches see http://www.informationisbeautiful.net/visualizations/worlds-biggest-data-breaches-hacks/.

[iii] ANSI : American National Standard Institute. X9: Accredited Standard Committee for financial services.

[iv] Enigma whitepaper.